\documentclass[aps,prl,twocolumn]{revtex4-1}

\usepackage{graphicx}
\usepackage{amsfonts}
\usepackage{amsmath}
\usepackage{color}
\usepackage{bm}
\usepackage{amssymb}

\usepackage{multirow}
\usepackage{epstopdf}

\usepackage{IEEEtrantools}

\begin{document}

\title{Effects of long-range hopping and interactions on quantum walk in ordered and disordered lattices}

\author{T. Chattaraj}
\author{R. V. Krems}%
\affiliation{%
University of British Columbia\\
 Vancouver, B.C., V6T 1Z1, Canada
}%


\date{\today}

\begin{abstract}
We study the effects of long-range hopping and long-range inter-particle interactions on quantum walk of hard-core bosons in ideal and disordered one-dimensional lattices.
We find that the range of hopping has a much more significant effect on the particle correlation dynamics than the range of interactions. We illustrate that long-range hopping makes the correlation diagrams asymmetric 
with respect to the sign of the interaction and examine the relative role of repulsive and attractive interactions on the dynamics of scattering by isolated impurities and Anderson localization in disordered lattices. 
We show that weakly repulsive interactions increase the probability of tunnelling through isolated impurities and decrease the localization. 

\begin{description}
%
\item[PACS numbers] 05.60.Gg, 71.23.An,  71.35.-y, 71.45.Gm
\end{description}
\end{abstract}

\pacs{Valid PACS appear here}
\maketitle


\section{\label{sec:level1}INTRODUCTION\protect\\  \lowercase{}}

Classical random walks of particles on a discrete lattice have many applications as algorithmic tools in computer science \cite{Lovasz, Adamic, Noh}.
The efficiency of many algorithms using random walks is determined by the time the walker takes to reach a stationary distribution (mixing time) and/or travel between two vertices on a graph (hitting time). 
The quantum dynamics of particles placed in individual sites of a lattice potential leads to quantum walks. Quantum walkers undergo interferences that may potentially accelerate the mixing and hitting times, improving the efficiency and scalability of the corresponding classical algorithms.  
This has stimulated the development of experiments to study quantum walks in lattice potentials. Quantum walks have been studied with photons in disordered media \cite{Lahini1, Silberberg0, Crespi, Silberberg1}, ions in rf traps \cite{Schaetz0, Feldker, Roos1} and ultracold atoms in optical lattices \cite{Karski, Billy, Bloch2, Greiner2}. The study of quantum walks in lattice potentials has potential to uncover novel features of quantum dynamics \cite{Kempe}, as evidenced by an abundance of theoretical work on quantum walks under various conditions \cite{Aharnov, Farhi, Vazirani, Ambainis}.

There is currently growing interest in developing experiments to study quantum walks with molecules trapped in optical lattices \cite{Weidemuller0, Danzl, Ospelkaus, Rempe, Bloch0, Sowinski, JunYe, Roman, Soderberg, Zoller1} or quasi-particles representing excitations of atoms or molecules trapped in optical lattices \cite{Roos2}. Molecules as well as some of the excitations \cite{Roos2, Weidemuller1, Weidemuller2, Rost, Sowinski, JunYe}  interact via long-range dipole - dipole interactions, which can be exploited to study the emergence of quantum correlations in quantum walks of interacting particles.  Understanding quantum correlations in few- and many-body dynamical systems is one of the central problems of current research in physics \cite{Bloch1, Greiner1, Bloch2, Greiner2}. Examples of interesting problems that can be probed in experiments with molecules or excitations in optical lattices include interaction-induced many-body localization (e.g., can many-body localization happen without disorder?) \cite{Schiulaz} or interaction-induced delocalization of quantum particles in disordered potentials \cite{Oppen, Shepelyansky, Imry, Shlyapnikov, Schreiber, Ortuno}. 

The experiments with excitations of molecules trapped in optical lattices are particularly appealing due to the unprecedented degree of control one can potentially achieve over the preparation and detection of quantum walkers \cite{Weidemuller0, Danzl, Ospelkaus, Rempe, Bloch0, Lang, Sowinski, JunYe}. In such systems, the molecules are fixed in individual lattice sites and quantum walkers are excitations analogous to wavepackets of  Frenkel excitons or magnons in molecular solids \cite{Gatti}. The excitations are resonantly transferred between molecules in different lattice sites due to long-range dipole - dipole interactions between the molecules. The excitations can be placed on individual molecules by applying a gradient of an electric field and resonantly exciting the corresponding molecules \cite{DeMille}. The excitations can be detected by applying a gradient of an electric field and resonantly ionizing the corresponding molecules \cite{Felipe0}. The study of quantum walk of Frenkel excitons in molecular ensembles in optical lattices can be used to elucidate the general features of quantum dynamics of excitations in molecular solids, ideal or doped. The properties of these excitations can also be mapped on the properties of polarons in both organic semiconductors and ionic solids \cite{Felipe0, Felipe1}. 

Most previous studies of quantum walks in lattice potentials have focussed on systems that permit direct transitions of particles only between adjacent lattice sites. However, the excitations of polar molecules in optical lattices, are characterized by the tunnelling amplitudes that decay as $t_{ij} \propto |\bm r_{ij}|^{-3}$, where $\bm r_{ij}$ is the distance between the sites $i$ and $j$, thus allowing for direct transitions between distant lattice sites. Quasi-particles with such tunnelling properties also arise in experiments probing excitonic energy transfer in molecular crystals \cite{Mukamel0}, molecular aggregates \cite{Fidder}, photo-sythetic complexes \cite{Fleming0, Fleming1, Linnanto, Mukamel1, Sarovar,  Engel0, Engel1}, artificial light-harvesting materials \cite{AspuruGuzik}, and ensembles of Rydberg atoms \cite{Weidemuller2, Weidemuller3}. The long-range tunnelling amplitudes are known to have a significant effect on the dynamical properties of quantum particles in lattice potentials \cite{Malyshev1, Malyshev2, Richerme, Pino, Hildner, Hauke, Adams}. In particular, the long-range tunnelling events may substantially decrease the mixing or hitting times of the corresponding quantum walkers. If so, quantum walkers with long-range hopping may be advantageous for quantum computing applications. 

Motivated by these experiments and the potential applications of quantum walkers, the present article studies quantum walk of two particles characterized by dipolar hopping $t_{ij} \propto |\bm r_{ij}|^{-3}$ and long-range inter-particle interactions $v_{ij}$ in ideal and disordered one-dimensional (1D) lattices. We explore the effect of the long-range character of each of $t_{ij}$ and $v_{ij}$ on the development of quantum correlations between the particles and the localization dynamics in disordered systems. We find that the long-range character of interactions $v_{ij}$ has a much smaller effect on the inter-particle correlations than the long-range character of $t_{ij}$ and that the long-range dependence of $t_{ij}$ on $r_{ij}$ introduces significant asymmetry in the two-particle correlation diagrams with respect to the sign of the inter-particle interaction. 

This study of quantum localization of interacting particles in disordered lattices builds on many previous articles debating the role of interactions\cite{Ortuno, Flach, Albrecht, Schreiber, Frahm, Shlyapnikov, Albrecht, Shepelyansky, Imry} on Anderson localization \cite{Anderson}. 
Following the work Lahini et al \cite{Lahini3, Lahini2} on quantun localization of photons, we approach the Anderson localization problem from the time-dependent perspective by 
examining the quantum walks of interacting pairs in disordered lattices in order to understand how interactions affect the localization dynamics.
The particular focus of the present work is on the interplay of the long-range tunnelling and interactions in the localization dynamics. We show that the $t_{ij}$-induced asymmetry in the two-particle correlations modifies significantly the particle - impurity interactions, leading to significant changes in the localization dynamics and attempt to correlate the conclusions with the results for the localization length of the interacting particles. Anderson localization of two interacting particles can be contrasted with many-body localization of particles coupled by long-range interactions \cite{Lukin, Hauke}, to learn about many-body effects driven by long-range interactions.

\section{Theoretical details}

We consider the quantum walk of two interacting hard-core bosons in a 1D lattice with $N$ sites described by the following Hamiltonian:
    \begin{equation}
     H =   \sum_{i} \sum_{j}   t _{ij}  a_i^\dagger  a_j + \sum_{i} \sum_j  v_{ij}  a_i^\dagger a_j^\dagger a_j a_i,
     \label{hamiltonian}
    \end{equation}  
where    $a_j$ is the operator that removes a particle from site $j$,
    \begin{eqnarray}
           t_{ij} & = & \frac{t}{|i-j|^\alpha}    
           \label{hopping}
    \end{eqnarray} 
is the hopping amplitude, and
    \begin{eqnarray}
           v_{ij} & = & \frac{v}{|i-j|^\beta}    
           \label{interaction}
    \end{eqnarray} 
is the strength of the interaction between particles in different lattice sites.     
The particle operators follow mixed statistics:
    \begin{eqnarray}
       a_i a_j^\dagger - a_j^\dagger a_i & = & 0,     \hspace{0.5cm} (i \neq j) \\
       a_i a_i^\dagger + a_i^\dagger a_i & = & 1, \label{hard-core} 
    \end{eqnarray}    
 Eq. (\ref{hard-core}) imposes the hard-core condition preventing two particles from occupying the same lattice site.    
We consider both attractive ($v/t < 0$) and repulsive ($v/t > 0$) interactions.

We note that an interaction $v(\bm r)$ is long-range if the integral $\int v(\bm r) d \bm r$ diverges. Thus, $v(\bm r) \propto 1/| \bm r| $ is long-range, while 
$v(\bm r) \propto 1/| \bm r |^3$ is short-range in 1D lattices. The long-range vs short-range character of hopping amplitudes is known to be critically important for dynamics of quantum particles in disordered systems. 
For example, non-interacting quantum particles with long-range hopping do not undergo Anderson localization in systems of infinite size \cite{Levitov0, Levitov1, Levitov2, Levitov3}.  Therefore, we distinguish dipolar hopping ($\alpha = 3$) and interactions ($\beta = 3$) from 
long-range hopping ($\alpha = 1$) and interactions ($\beta = 1$), and examine the dynamics of systems with all combinations of $\alpha$ and $\beta$ values. The results with $\alpha = 3$ and $\beta = 3$ describe quasi-particles corresponding to excitations confined in 1D lattices, while the results with $\alpha = 1$ and $\beta = 3$ reveal qualitative features expected of excitations in 3D systems.

We consider the dynamics of particles initially in well-defined sites $\tilde i$ and $\tilde j$ of a one-dimensional lattice. The time evolution of the quantum state 
$\Psi  (\tau = 0) =   | \tilde  i \tilde j \rangle$ is calculated using the eigenenergies $E_n$ and eigenstates $|n\rangle$ of the full Hamiltonian as follows: 
     \begin{equation}
     \Psi   ( \tau) = \sum_n  \exp{\left (-\frac{\it i \normalfont E_n \tau}{\hbar}\right)} \vert n \rangle \langle n \vert \Psi(0)\rangle 
     \label{wave-function}
    \end{equation}
Given the time-dependence of the full state, we compute the pair correlations (or joint probabilities) defined as
     \begin{equation}
      \Gamma_{ij} = \langle a_i^\dagger a_j^\dagger a_j a_i \rangle, 
     \end{equation}
and the particle density distributions in real space.

To explore quantum walk in disordered systems, we divide the lattice sites into two subsets $P$ and $Q$, with $p N$ sites in the $P$ subset and $(1-p)N$ in the $Q$ subset. For a given value $p$, the lattice sites are assigned to the subsets at random. The hopping amplitudes of the Hamiltonian (\ref{hamiltonian}) are then defined as follows:
\begin{align}
t_{ij} &= \begin{cases} 
      {t}/{|i - j|^{\alpha}}, & i \in P \text{ and } j  \in P \\
      0, & i \in Q \text{ and/or } j  \in Q
   \end{cases} \label{tijDef}
\end{align}
The impurities are thus represented by vacancies and the Hamiltonian (\ref{hamiltonian}) describes interacting particle in a disordered, diluted lattice with $pN$ sites, relevant for many experimentally realizable systems. For example, the Hamiltonian (\ref{hamiltonian}) with the hopping amplitudes (\ref{tijDef}) describes the dynamics of excitation transfer in a system of molecules trapped in a 1D optical lattice with $p$ empty sites \cite{JunYe, JunYe0, JunYe1, JunYe3} or in a molecular crystal with vacancies.

  To characterize the localization in disordered systems, we use the wave function (\ref{wave-function}) to compute the  participation ratio ($P$) defined as follows:
      \begin{equation}
         P^{-1} =  \sum_i \Psi_{i}^{2}(\tau),
    \end{equation}
where    

\begin{eqnarray}
\Psi_i = \frac{1}{2} \sum_{j \neq i} \Psi_{ij}^\dagger \Psi_{ij}
\end{eqnarray}
 and $\Psi_ {ij} = \langle i j | \Psi \rangle$.  

We compute  the participation ratio at a long time $\tau$ greatly exceeding the Heisenberg time and average $P$ over multiple realizations of disorder to obtain $\langle P \rangle$. 
  For particles undergoing localization, the quantity $\langle P \rangle$ in the limit of long time is time-independent and 
 inversely proportional to the localization length. 

\section{Results}

\subsection{Quantum walk in ideal lattices}

Figures 1 and 2 present the pair correlations comparing the results for different values of $\alpha$ and $\beta$ in Eqs. (\ref{hopping}) and (\ref{interaction}). The values of the pair correlation function along the diagonal describe particles walking in concert, while the values in the upper 
left and lower right corners of the diagrams give the probability of anti-walking. The examination of Figures 1 and 2 reveals the following interesting observations: 

\begin{itemize}

\item Extending the range of hopping $t_{ij}$ changes the correlation dynamics significantly, while extending the range of interactions $v_{ij}$ has little to no effect on the correlation dynamics.

\item When the amplitudes $t_{ij}$ allow hopping between nearest neighbour (NN) sites only, the correlation diagrams are symmetric with respect to the sign of the interaction ($\pm v/t$). This symmetry is broken when the hopping is extended beyond NN sites. 
This asymmetry is more significant for smaller values of $\alpha$. 

\item When the interactions are strong, both the repulsive and attractive interactions increase the probability of co-walking. 

\end{itemize}

These observations can be rationalized by the analysis of the dispersions of non-interacting particles shown in Figure 3 for NN hopping and different values of $\alpha$. The band of the two-particle continuum states for pairs of particles with NN hopping 
is symmetric with respect to the zero energy axis. This symmetry manifests itself in the symmetry of the pair correlation diagrams with respect to the sign of the interaction. The interaction magnitude $|v/t| = 2$ marks the onset of a bound state that splits from the continuum at negative energies for $v/t < - 2$ and at positive energies for $v/t > 2$. The bound state splitting to above the continuum is sometimes referred to as the repulsively bound state \cite{Zoller2}. The wave functions of the attractively and repulsively bound states are identical \cite{Ping}. As the interaction strength increases, the contribution of the bound state to the wave packet of the walking pair increases, leading to enhanced probability of co-walking. 

When the hopping range is extended beyond nearest neighbours, the dispersion bands lose their symmetry (see Figure 3). The distortion decreases the energy required to form an attractively bound state and increases the energy to the upper edge of the continuum band. It thus requires less energy to form a bound state at negative energies than to split a state from above the continuum. In the limit of infinitely large lattices, the upper and lower bounds for the lattice spectra can be calculated from the values of Riemann zeta 
functions and are (in the units of $t$) equal to  ($+4.80, -3.60$), ($+6.58, -3.30$), ($+\infty, -2.80$) respectively for  $\alpha = 3, 2$ and $1$.  

Figures 1 and 2 also show that in the absence of interactions ($v = 0$), the quantum walk of pairs with NN and dipolar hopping are similar. The correlation diagrams in these cases illustrate a strong propensity for particles to move in the opposite directions, which is equivalent to anti-bunching of fermions (cf., Fig 3 in Ref. \cite{Bordone}). In the case of long-range hopping ($\alpha = 1$), the anti-bunching of hard-core bosons is much less pronounced. One can observe in the lower middle panel of Figure 1 a non-zero probability of particles to remain in adjacent lattice sites as they explore the lattice. These correlations are, however, significantly suppressed and the quantum walk, while less dominated by the anti-walking correlations, exhibits the propensity of the particles to stay apart.

\subsection{Scattering by isolated impurities}

The dynamical properties of non-interacting particles in disordered lattices are well understood. 
For particles with short-range hopping ($\alpha \le 1$ in Eq. (\ref{hopping}) for 1D lattices), scattering by a random distribution of impurities and/or lattice site vacancies leads to Anderson localization, which precludes the quantum transport. 
The effects of particle interactions on Anderson localization have been the subject of many recent studies \cite{Ortuno, Albrecht, Schreiber, Frahm, Shlyapnikov, Albrecht, Shepelyansky, Imry}. However, there are still many questions that remain open. In particular, the effect of interactions on scattering by an isolated impurity, important for the sub-diffusive and diffusive regimes of transport in disordered lattices with a small concentration of impurities, has not been considered. 

In order to examine the interaction of correlated pairs in the limit of isolated impurity (as opposed to multiple impurity) scattering, we compute the quantum walk of two particles placed in adjacent lattice sites ($i=20;  j=21$) in the middle of a lattice with two impurity sites  at $k=10$ and $l = 31$. To model the interactions of excitons in molecular crystals with lattice vacancies, we introduce the impurities by setting $t_{i,k=10}$
 and $t_{i,l=31}$  to zero for all values of $i$. 
The impurities thus act as effective potential barriers localized to a single site. The extended range of the hoping amplitudes allows tunnelling under the scattering barriers. In order to eliminate the finite-size effects leading to reflection of wave packets from the lattice boundaries, we impose the absorbing boundaries outside the vacant sites that annihilate the fraction of the wave packet that tunnels under the barriers. Figures 4 and 5 show the fraction of the particle density that remains at time $20/t$ between the impurities as a function of the interaction strength. 

The comparison of Figures 4 and 5 reveals interesting details of the effect of interactions on particle - impurity scattering dynamics. 
As can be seen from Figures 4 and 5, both strongly attractive and strongly repulsive interactions reduce the tunnelling probability. The range of the interaction has a relatively small effect on the particle - impurity scattering dynamics. On the other hand, changing the range of hopping for the same range of interactions modifies the tunnelling dynamics to a great extent. For particles with dipolar hopping, the tunnelling probability passes through a maximum when the particle - particle interactions vanish. The interaction-induced particle - particle correlations thus generally suppress the tunnelling of particles with dipolar hopping under impurity barriers. In the case of long-range hopping (\ref{hopping}) with $\alpha = 1$, the correlations induced by repulsive interactions with weak to moderate strength enhance the tunnelling probability, leading to a minimum of the survival probability at $v/t \approx 5$ (see Figure 5).

The decrease of the tunnelling probability in the presence of both attractive and strongly repulsive interactions reflects the contribution of the repulsively and attractively bound states in the wave packet dynamics. Binding increases the effective mass of the particles, leading to slower tunnelling. On the other hand, repulsive interactions increase the kinetic energy of the separating particles, which should enhance the tunnelling probability. For particles with dipolar hopping,  the former is the dominant effect for repulsive interactions with $v/t > 2$. 
At the interaction strengths $v/t = 0 - 2$, the two effects appear to balance each other. 
For particles with $\alpha =1$, the dispersion spectrum is distorted so significantly that the effect of the repulsive interactions on the kinetic energy is more dominant for interactions with $v/t = 0 - 5$. 

\subsection{Localizaton in a random potential}

In order to explore the role of particle interactions on quantum localization in a random potential, we examine quantum walk of interacting particles in a 1D lattice with a random distribution of impurities.  
As described in Section II, the impurities are introduced by prohibiting access to certain lattice sites, drawn from a random distribution. This leads to scattering and tunnelling dynamics as discussed in the previous sub-section, but in the regime of multiple-impurity scattering.   
For 1D lattices, the scattering dynamics of particles with dipolar hopping in the presence of a random distribution of impurities must lead to Anderson localization.  
While particles with $1/r_{ij}$ hopping ($\alpha = 1$) are not expected to be localized in 1D lattices of infinite size \cite{Levitov0, Levitov1, Levitov2, Levitov3}, they may undergo localization in lattices of finite size \cite{Josh}. 
For comparison purposes, we perform calculations for both $\alpha =1$ and $\alpha = 3$.

Figure 6 presents the correlation diagrams for particles with long-range hopping ($\alpha =1$) and dipolar hopping ($\alpha = 3$) undergoing quantum walk in disordered 1D lattices. The results are averaged over $5000$ realizations of disorder and presented for the case of dipolar interactions $\beta = 3$. 
As can be seen in the left column of Figure 6, disorder completely suppresses anti-bunching and the quantum walk is dominated by particles moving together, even at zero inter-particle interactions. 
We find that, as expected, any degree of attractive interactions further enhances the correlations shown in the left column of Figure 6.  Surprisingly, Figure 6 illustrates that repulsive interactions, even relatively weak with magnitudes $v/t \approx 2$, also lead to strong correlations forcing the particles to walk in concert. This is particularly evident in the comparison of the upper right panel of Figure 6 with the lower right panel of Figure 1. 
In the case of dipolar tunnelling, even a weakly repulsive interaction with $v/t=1$ is sufficient to induce strong correlations, effectively binding the particles in disordered lattices.

The comparison of the correlation diagrams with $v = 0$ and $v/t > 0$ in Figure 6 reveals that weakly repulsive interactions extend the range of the lattice explored by the particles, effectively making the particles less localized. 
To examine the role of interactions on the localization length, we compute the participation ratio as a function of the interaction strength (see Figure 7). For particles with dipolar hopping, the participation ratio as a function of the interaction strength exhibits a maximum between $v/t=1$ and $v/t=2$. This indicates that weakly repulsive interactions decrease the localization, while attractive as well as strongly repulsive interactions enhance the localization. This is consistent with the results for tunnelling through impurities discussed in the previous sub-section (cf., Figures 4 and 5).

We note that the results in Figure 6 are presented for a fixed time greatly exceeding the Heizenberg time of this system. 
By computing the correlation diagrams at several different times, we have verified that, after averaging over $5000$ disorder realizations, the results in Figure 6 are time-independent. 
While our tests do not preclude diffusion of the wave packets to the lattice edges at infinite time, Figure 6 illustrates localization at finite times relevant for experiments.  This should be considered in the context of the results of Ref. \cite{Flach} suggesting that weakly repulsive interactions bring localized particles in lattices of infinite-size to a sub-diffusive regime.

\section{Conclusion}

In the present work, we examine the role of interactions and the range of hopping amplitudes on quantum walk of hard-core bosons in one-dimensional lattices. 
We have demonstrated that, unlike for systems with the kinetic energy described by the tight-binding model, for particles with hopping extended beyond nearest neighbour sites the effects of repulsive and attractive interactions are different. 
Our results illustrate the effects of interactions on particle - impurity scattering and particle mobility in disordered systems. In particular, we have found that weakly repulsive interactions enhance the probability of tunnelling through impurities in 
one-dimensional systems and decrease the localization of particles in disordered lattices. At the same time, the localization increases in the limit of strong interactions, whether repulsive or attractive. This raises the question of whether 
particles with dipolar hopping may undergo an interaction-induced transition to the diffusive regime in two-dimensional disordered lattices.

The results presented here can be directly tested 
in experiments on rotational excitations of ultracold molecules \cite{JunYe, JunYe0, JunYe1, JunYe3} or Zeeman excitations of highly magnetic atoms \cite{Zoller3} trapped in optical lattices. In these experiments, the rotational excitations or the Zeeman excitations are hard-core bosons that undergo transitions between atoms or molecules trapped in different sites of an optical lattice due to the dipole - dipole interactions. In the case of polar molecules, the interactions 
between the rotational excitations can be tuned by an external electric field \cite{Ping}. The relative magnitude of the hopping amplitude and the interactions between the Zeeman excitations of highly magnetic atoms can be tuned by transferring atoms into different Zeeman states, as shown in Ref. \cite{Rodrigo}.

The Hamiltonian (\ref{hamiltonian}) with  long-range hopping (\ref{hopping}) describes excitons in molecular crystals \cite{Logan}, photo-sythetic complexes \cite{Fleming0, Fleming1, Linnanto, Mukamel1, Sarovar,  Engel0, Engel1}, $J$-aggregates \cite{Kobayashi, Yefimova, Fang}, and ensembles of Rydberg atoms \cite{Zoller0, Weidemuller1, Lesanovsky, Secker} in the absence of coupling to their phonon environment. The present work indicates that a proper description of exciton correlations in these systems must account for long-range hopping of excitations unless quantum interferences are destroyed by the environment.




\newpage
\begin{figure*}[p]
\includegraphics[width = 1.0\textwidth]{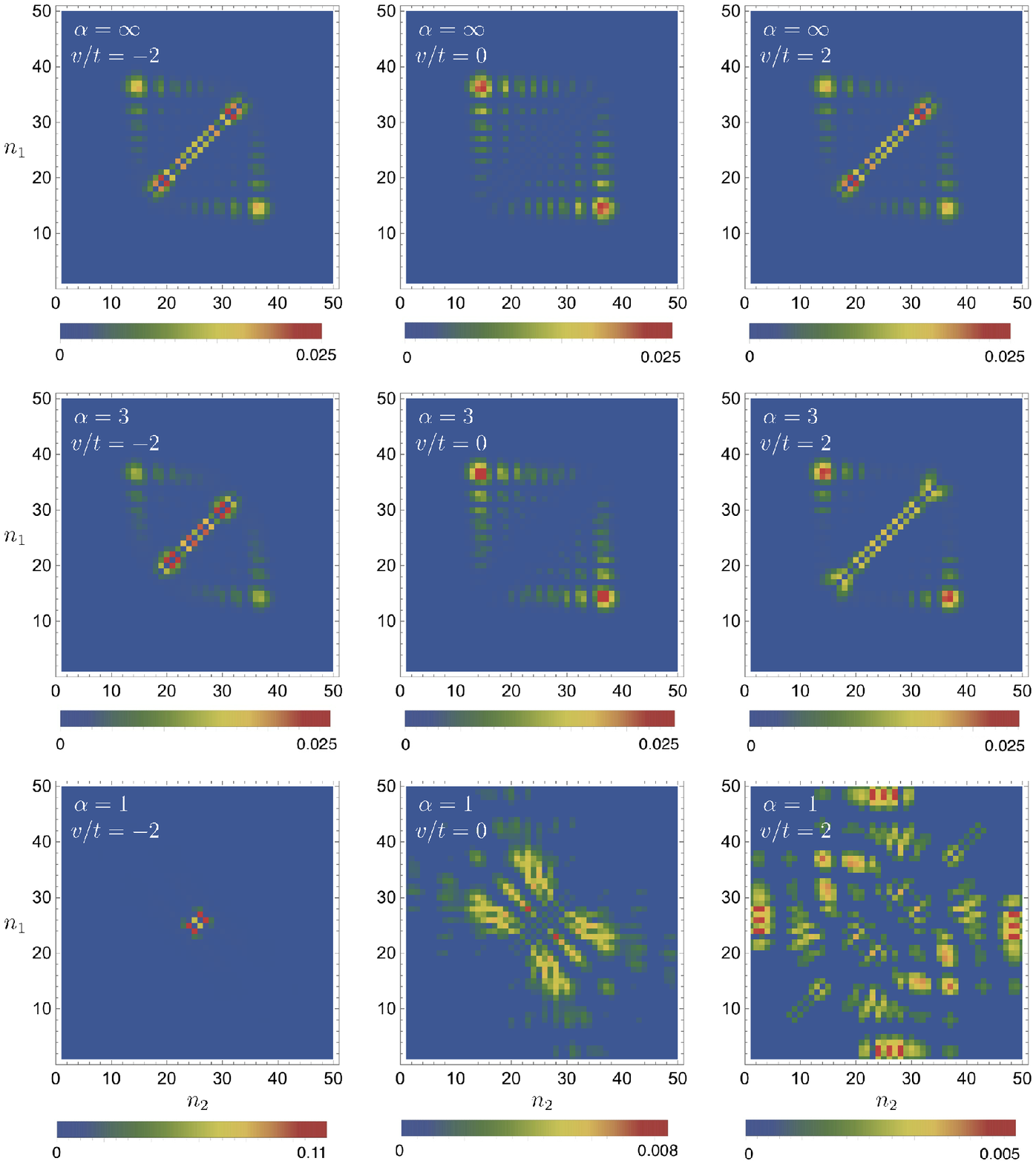}
\caption{Pair correlations for two hard-core bosons with $\beta = 1$ in Eq. (\ref{interaction}) computed at time $2\pi / t$ in a 1D lattice with 50 sites. The particles are initially in sites $i = 25$ and $j = 26$. The values along the diagonal are the probabilities of co-walking. The labels $n_1$ and $n_2$ denote the site indexes of particles 1 and 2, respectively.}
\end{figure*}

\newpage
\begin{figure*}
\includegraphics[width = 1.0\textwidth]{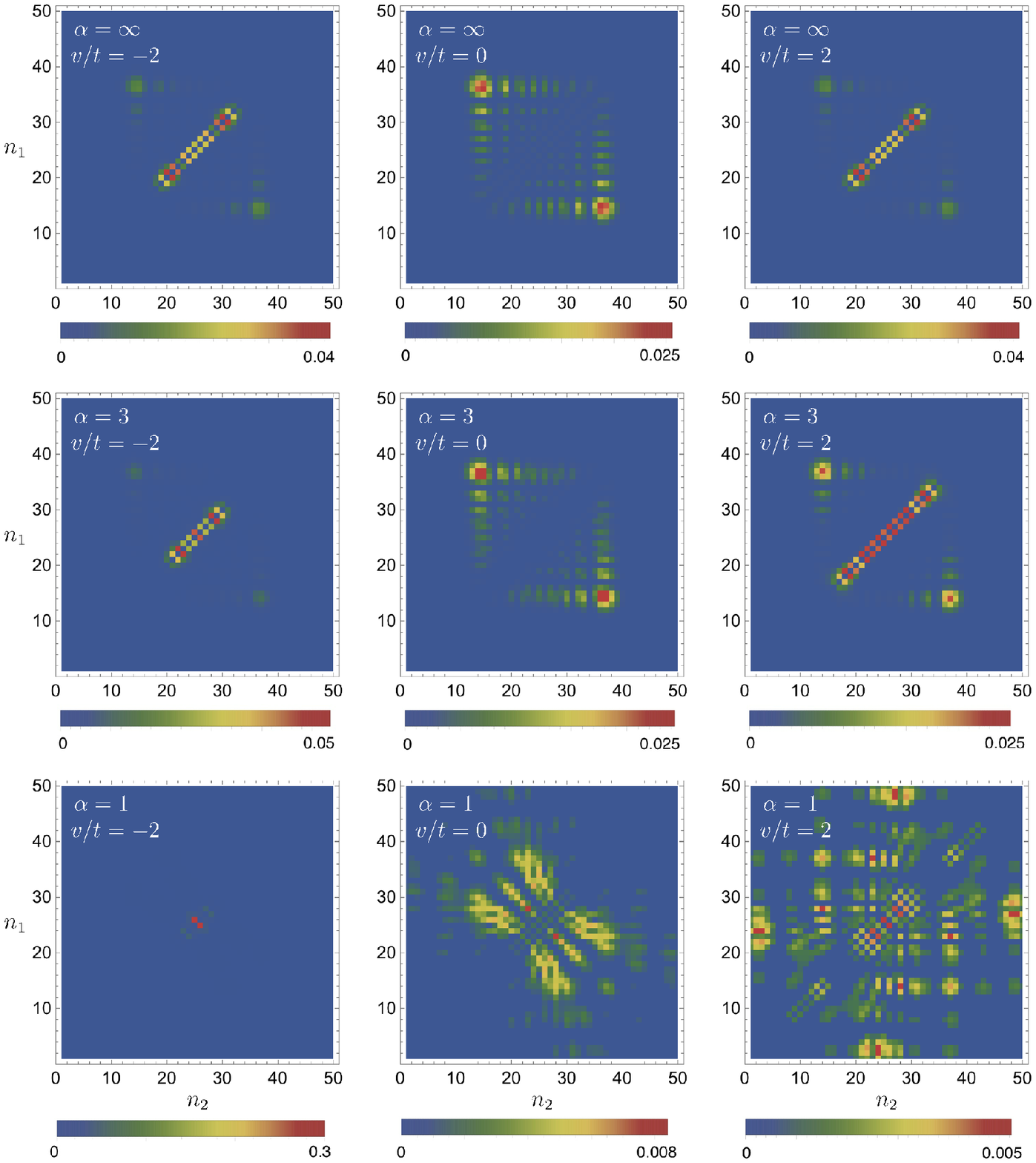}
\caption{Pair correlations for two hard-core bosons with $\beta = 3$ in Eq. (\ref{interaction}) computed at time $2\pi / t$ in a 1D lattice with 50 sites. The particles are initially in sites $i = 25$ and $j = 26$.  The labels $n_1$ and $n_2$ denote the site indexes of particles 1 and 2, respectively.}
\end{figure*}

\newpage
\begin{figure*}
\includegraphics[width = 1.0\textwidth]{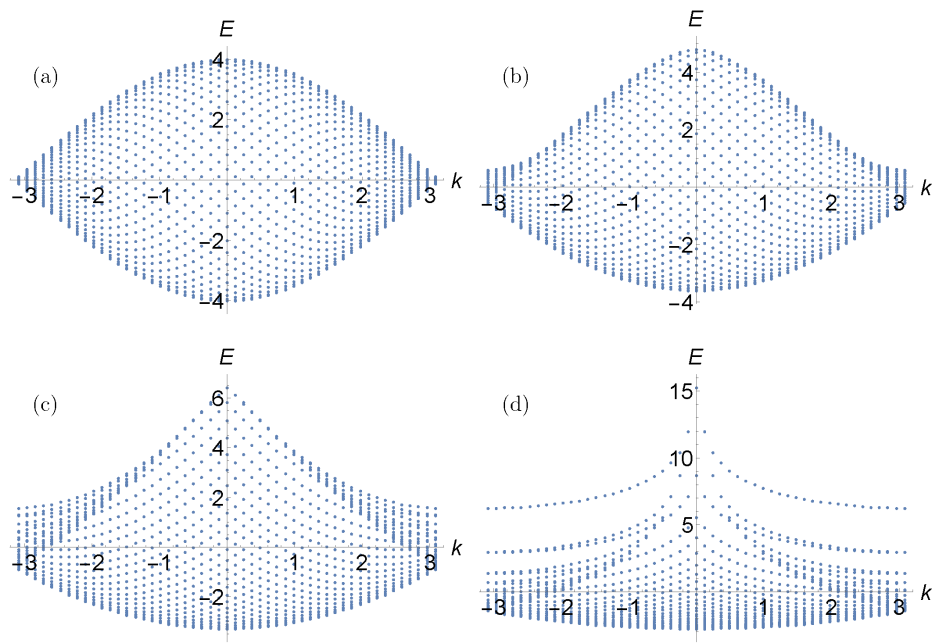}
\caption{Eigenstates of two non-interacting hard-core bosons in a 1D lattice. The different panels correspond to the different range of hopping $t_{ij}$: (a) nearest neighbour hopping, (b) $\alpha = 3$, (c) $\alpha = 2$, (d) $\alpha = 1$.
The energy is in the units of $t$ and wave number -- in the units of $1/a$, where $a$ is the lattice constant. 
}
\end{figure*}

\newpage
\begin{figure*}
\includegraphics[width = 1.0\textwidth]{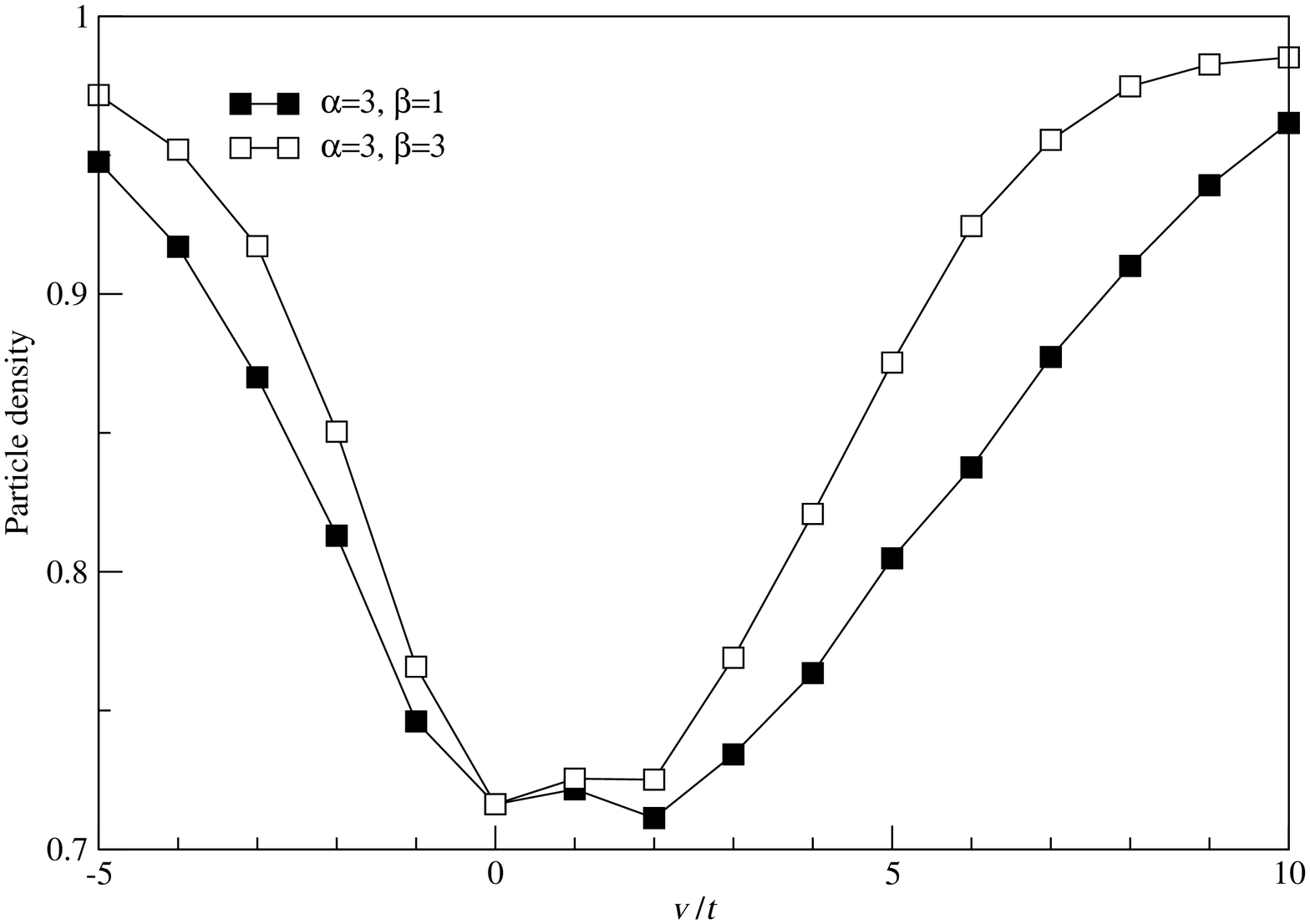}
\caption{Particle density $\cal P$ integrated over the part of the lattice between two impurities at sites 10 and 31 at time $20/t$ for two particles with long-range hopping and long-range interactions initially placed in a 1D lattice at sites 20 and 21. The part of the wave packet tunnelling to $i < 10$ and $i> 31$ is absorbed to eliminate reflection from the lattice boundaries, thus leading to irreversible decay of $\cal P$ as a function of time. The values of $\alpha$ and $\beta$ specify the parameters of hopping (\ref{hopping}) and interactions (\ref{interaction}).}
\end{figure*}

\newpage
\begin{figure*}
\includegraphics[width = 1.0\textwidth]{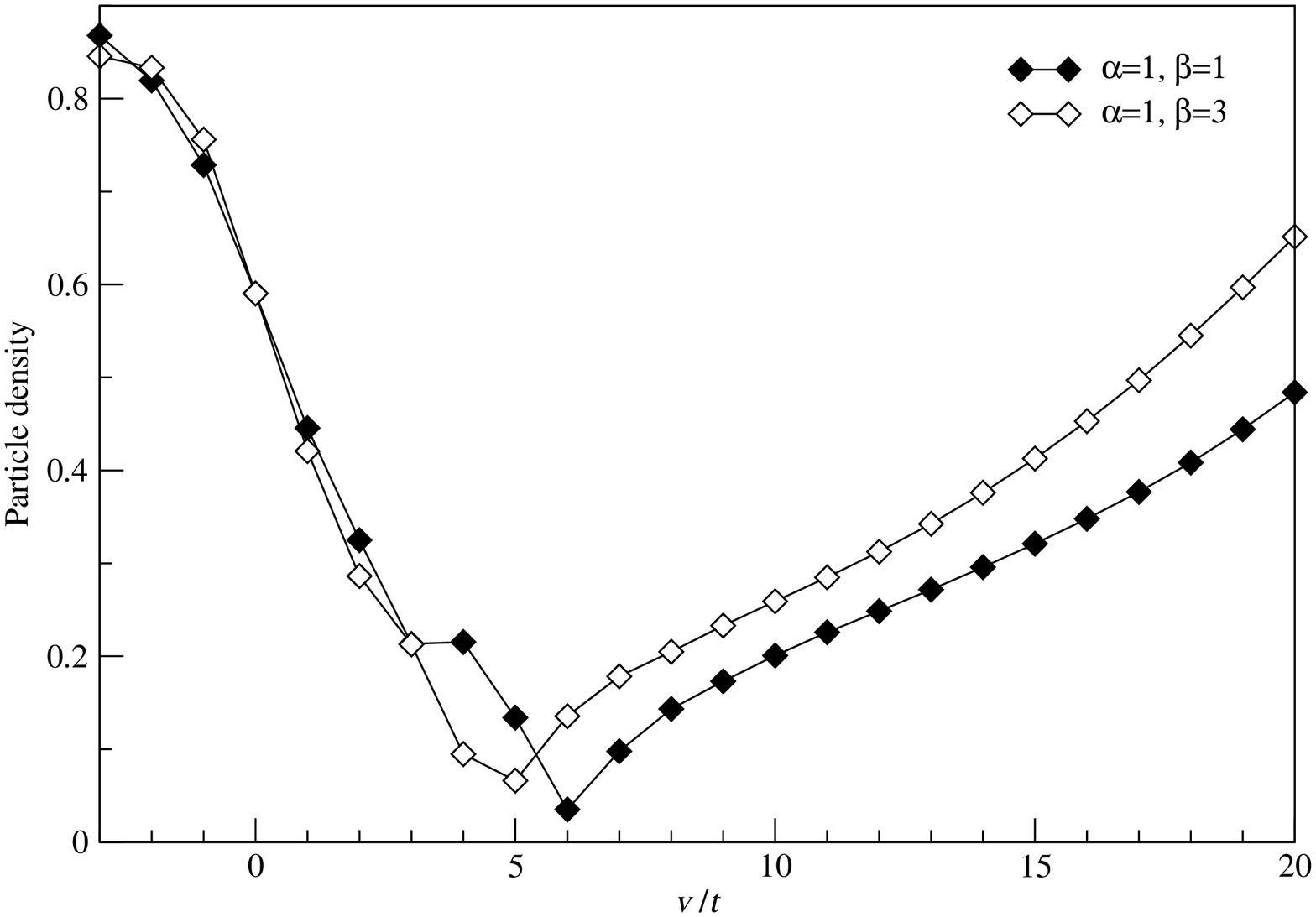}
\caption{Particle density $\cal P$ integrated over the part of the lattice between two impurities at sites 10 and 31 at time $20/t$ for two particles with long-range hopping and long-range interactions initially placed in a 1D lattice at sites 20 and 21. The part of the wave packet tunnelling to $i < 10$ and $i> 31$ is absorbed to eliminate reflection from the lattice boundaries, thus leading to irreversible decay of $\cal P$ as a function of time. The values of $\alpha$ and $\beta$ specify the parameters of hopping (\ref{hopping}) and interactions (\ref{interaction}).}
\end{figure*}

\newpage
\begin{figure*}
\includegraphics[width = 1.0\textwidth]{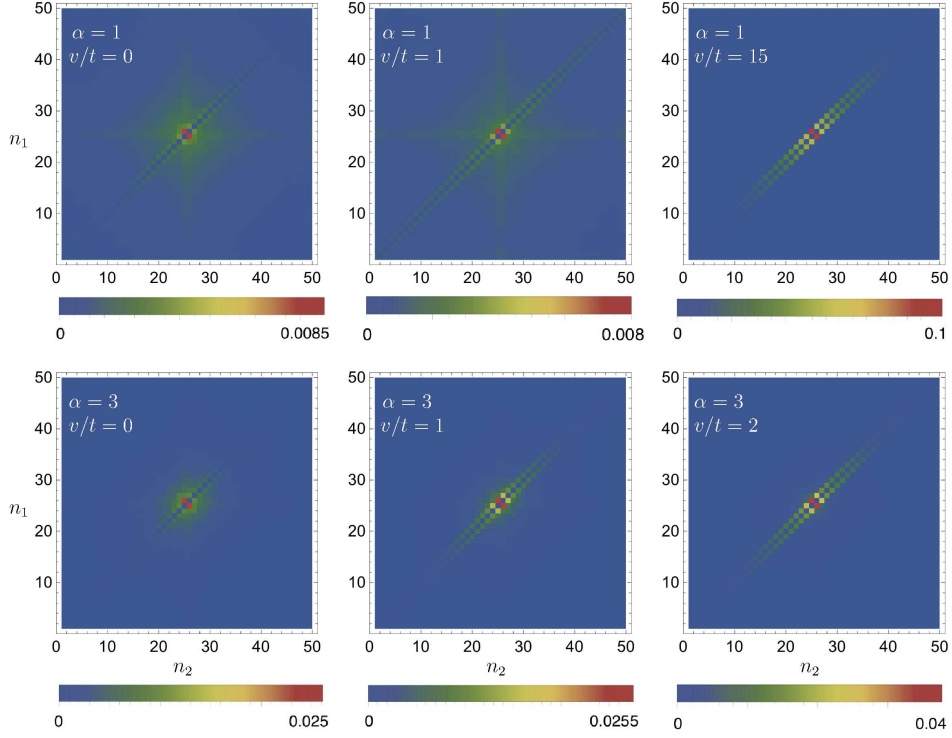}
\caption{Pair correlations in the limit of long time $\tau = 10^4/t$ for two hard-core bosons with long-range hopping and dipolar interactions $\beta = 3$ in a 1D lattice 
disordered by a random distribution of impurities with the concentration $p = 0.1$. The results are averaged over $5000$ realizations of disorder. The labels $n_1$ and $n_2$ denote the site indexes of particles 1 and 2, respectively.}
\end{figure*}

\newpage
\begin{figure*}
\includegraphics[width = 1.0\textwidth]{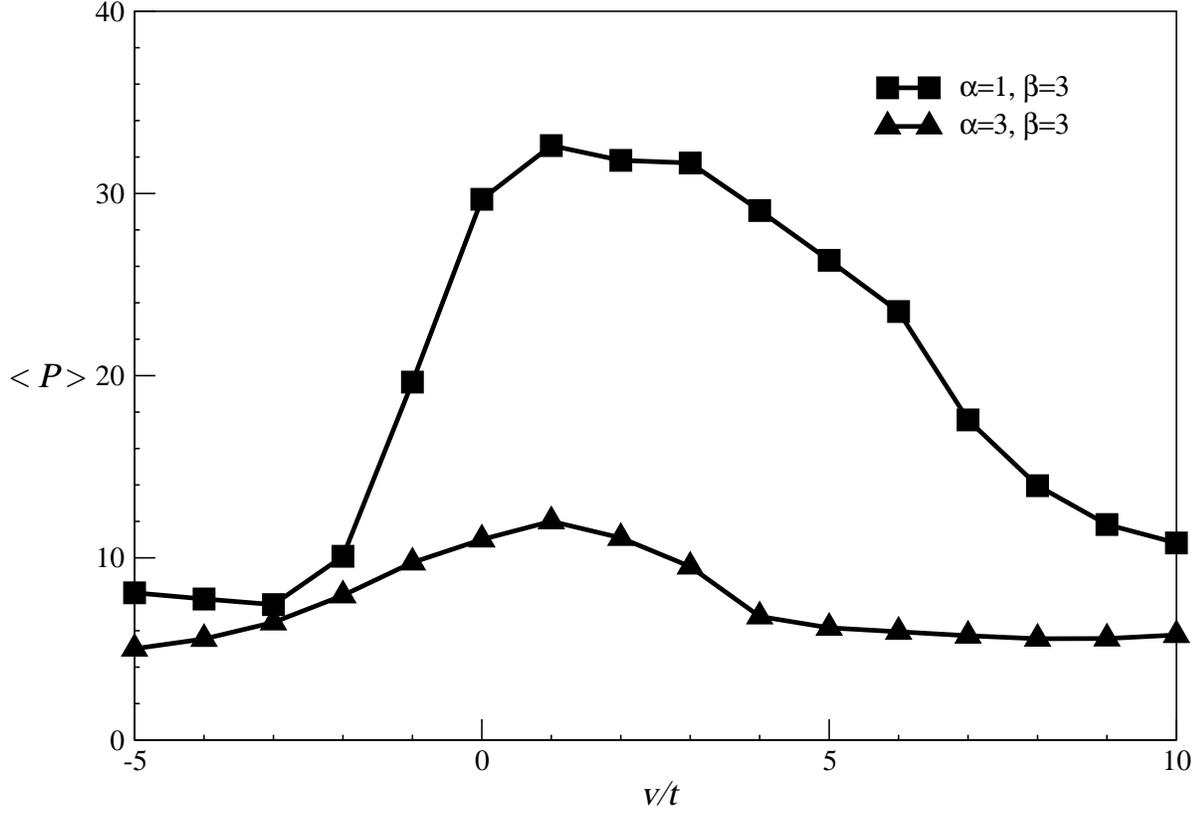}
\caption{The inverse participation ratio $\langle P \rangle$ calculated at $ \tau = 10^4/t$ from localized density distributions for two particles initially placed in adjacent sites of a disordered 1D lattice with $N = 50$ and $p = 0.9$. The results are averaged over $5000$ realizations of disorder.}
\end{figure*}

\newpage

\end{document}